\documentclass[12pt, a4paper]{article}
\usepackage{arxiv}

\usepackage{cmap}
\usepackage[british]{babel}

\usepackage{amsmath}
\usepackage[OT1,euler-digits,euler-hat-accent]{eulervm}  

\usepackage{graphicx}
\graphicspath{ {./} }

\usepackage{enumitem}
\setlist{nolistsep}

\usepackage{hyperref,bookmark}
\usepackage{listings}
\usepackage{subcaption}
\usepackage{cite}
\usepackage{multirow}

\makeatletter
\newenvironment{wraptable}
  {\def\@captype{table}}
  {}

\newenvironment{wrapfigure}
  {\def\@captype{figure}}
  {}
\makeatother

\title{An approach to measure route quality and refine the route during the voyage using characteristic coefficients}

\author{
  Ivan Yanchin \\
  Department of Computer Sciences \\
  State Marine Technical University \\
  Saint-Petersburg\\
  Russia \\
  \And
  Oleg Petrov \\
  Department of Computer Sciences \\
  State Marine Technical University\\
  Saint-Petersburg\\
  Russia \\
}

\date{}

\renewcommand{\headeright}{}
\renewcommand{\undertitle}{}

\usepackage{multicol,caption}

\begin{document}

\maketitle

\begin{abstract}
The paper presents a method to validate and refine the ship's route during the voyage. 
The method is based on computing several characteristic coefficients that represent and measure route properties. Thru the analysis of the values of these coefficient, one can analyse the overall route quality and detect possibly dangerous discrepancies between the actual route and the planned route.The paper describes the proposed characteristic coefficients, the process of route refinement and the method for prediction and validation of the route's future changes.

\textbf{keywords}: marine navigation; marine safety; routing; route refinement.
\end{abstract}

\begin{multicols}{2}
\section{Introduction}
Planning safe and optimal route for a ship is known to be an important and difficult task~\cite{2015-simonsen-state-of-the-art-weather-routing, 2018-grifoll-ship-weather-pathfinding}. It has been demonstrated that if weather conditions are taken into account, the resulting route may be shorter in terms of time and cheaper in terms of costs even if it is longer in terms of distance~\cite{2018-grifoll-weather-ship-routing-economics}.

In case of navigation safety, it may be more important not just to plan a safe route, but to detect that the actual route deviates from the planned safe one. According to the official report on the investigation of the Costa Concordia's case ``the ship was sailing too close to the coastline, in a poorly lit shore area, ... at an unsafe distance at night time and at high speed ... the danger was considered so late that the attempt to avoid the grounding was useless''~\cite{2013-mit-costa-concordia-report}. According to the annual overview of marine casualties and incidents of the 2018 year issued by European Maritime Safety Agency, up to 58\% of all maritime incidents are caused by human factor and almost half of the incidents are navigational incidents (grounding, stranding or collisions)~\cite{2018-emsa-marine-casualties}. It is also noted, that most of the incidents have happened in port areas and internal waters~\cite{2018-emsa-marine-casualties}, which means, that the most dangerous areas are areas with relatively shallow waters and heavy traffic. Therefore, development and deployment of a system that analyses long-term results of human actions and decisions on the ship's route and its safety may help reduce the number of such navigational incidents.

Autonomous ships is an emerging technology that would change the way of shipping~\cite{2019-nikitakos-4-industrial-revolution-marine, 2018-benson-autonomous-transport, 2016-rr-marine-autonomous-ships}. One of the important concepts related to autonomous ships is the technologies used to plan safe and optimal route automatically, without any human intervention~\cite{2016-munin-final-brochure}. Such systems require the on-board route planning system to be able to estimate the route's safety and optimality, automatically, using formal criteria instead of the experience of the crew. Without the crew onboard, the role of the human factor decreases and the impact into the accidents of the technologies that power the ship becomes more important~\cite{2016-blanke-autonomous-ships, 2017-cross-autonomous-ships-101}. Therefore more attention is to be payed to the safety of the commads ordered by such on-board control systems.

The method proposed in this paper consists in the detection of deviations of the actual route from the planned one and analysis of the effects of these deviations on the route safety. Here, route deviation is the difference between a waypoint of an actual route and the corresponding waypoint of the planned route. Points can differ in terms of location and ETA and ETD, since these properties cause differences in other properties of a waypoint, like speed or movement direction. A deviation may be caused by effects of the environment, e.g. effects of wind waves or unexpected manoeuvre of another ship nearby, or by human action (or inaction, i.e. the crew may miss a waypoint). In addition, the safety of a route may be affected by the weather conditions (e.g., storms) that have not been expected or properly taken into account while planning the route. Since weather forecasts are used while planning a route, the actual weather conditions may differ from the forecast. The forecast may be refined during the voyage several times. After the updated information about the weather conditions is been received, it is important to ensure that the route has not become unsafe.

Within this paper, a route is considered to be a sequence of waypoints connected with edges. Each waypoint describes the manoeuvre the ship must perform to arrive at the next waypoint. Therefore, a waypoint describes the course angle the ship must have while following the next edge (the edge that starts at this waypoint), the speed it must move with and the distance it must pass. A waypoint is also associated with ETA and ETD. We assume that the departure time may differ from the arrival time, in this case, the ship must wait. Within this paper, we consider a ship that follows a particular route, however, the actual route deviates from the one planned initially. The task is to analyse whether these deviations are potentially dangerous for the ship in the long run in case the ship does not resolve them and does not return to the planned route.

This paper is organised as follows. Section 2 describes the way to measure the route's properties. Section 3 is dedicated to the method to predict the future route structure and examine it using the coefficients. Section 4 describes the method to analyze route development over time. Section 5 demonstrates the proposed method through the analysis of a real navigational accident. Section 6 concludes the paper and provides our findings.

\section{Characteristic coefficients of a route}

To analyze route safety and optimality, it is important to define criteria used to measure these properties and to compare different routes to determine which of them is better.

We propose to introduce four characteristic coefficients, each of which measures a single aspect of a route. These coefficients measure to what extent the route is close to the best possible route. The best possible route is a straight line that connects the start and the destination points. The ship follows this route with the optimal speed that it can achieve in calm waters.

A route needs to be safe for the ship, i.e. it must not make the ship crash into an obstacle. Therefore, the safety coefficient can be defined as 
\[
	S_r = 1 - P_r(collide)
\]
where $P_r(collide)$ is the probability of a ship to collide with an obstacle while following the route $r$.

As we have already said, the best route is a straight line that connects start and destination. If the best route is blocked by an obstacle, then the actual route is not a straight line. The distance coefficient measures to what extent the route is close to the best one in terms of distance. This coefficient is defined as 
\[
D_r = \frac{l_e}{l_r}
\]
where $l_e$ is the length of the best route (the length of the line segment between start and destination points) and $l_r$ is the length of the route being examined. The closer the route to the straight line, the greater the value of $D$ is. By definition, $0 < D \leq 1$.

The less time it takes to travel by a route the better the route is. The best possible route makes the ship travel by a straight line with the optimal speed, which means, that this route is optimal in terms of time. Therefore, the time coefficient is defined as

\begin{align*}
T(r) = \frac{t_e}{t_r}
	 &= \frac{\frac{l_e}{v_{max}}}{\sum_{i=1}^n [\frac{l_{r}^{(i)}}{v_{r}^{(i)}} + (t_{r,dep}^{(i)} - t_{r,arr}^{(i)})]} \\
	 &= \frac{l_e \sum_{i=1}^n [\frac{l_{r}^{(i)}}{v_{r}^{(i)}} + (t_{r,dep}^{(i)} - t_{r,arr}^{(i)})]}{v_{max}}
\end{align*}

where $t_e$ is the time needed to follow the best possible route, $t_r$ is the time needed to travel by the route under consideration, $l_e$ is the length of the best possible route, $v_{max}$ is the maximum possible speed of the ship in calm waters, $l_{r}^{(i)}$ is the length of the $i$-th edge of the route $r$, $v_{r}^{(i)}$ is the planned speed for the $i$-th edge of the route $r$, $t_{r,dep}^{(i)}$ is the planned departure time (ETD) for the $i$-th waypoint of the route $r$ and $t_{r,arr}^{(i)}$ is the planned arrival time (ETA) for the $i$-th waypoint of the route $r$. However, a ship can move faster than in calm waters in case of co-directed streams. Therefore, by definition and using this remark, $0 < T(r)$.

Paper~\cite{2019-jeong-multi-criteria-route-planning} emphasizes that better route prescribes fewer manoeuvres, therefore we define the simplicity coefficient as 
\[
C_r = \frac{2}{p_r}
\]
where $p_r$ is the number of waypoints the route $r$ has. The best route contains only two waypoints: the start point and the destination points, therefore, the coefficient $C$ has greater value for routes that have fewer waypoints and has a value of $1$ for the routes that have exactly $2$ waypoints. By definition, $0 < C \leq 1$.

According to the respective definitions, for the best possible route, all coefficients are $1$, and the time coefficient is equal to or greater than $1$. Therefore, for the best possible route,
\[
S_r + D_r + T_r + C_r \geq 4.
\]

Together, distance, time and simplicity coefficients form the value of the optimality criterion. Safety coefficient forms the value of the safety criterion. It is possible to extend the safety criterion with other coefficients that check, for example, whether there is a risk of broaching or slamming or extend the optimality criterion to take into account fuel consumption, which is considered important~\cite{2009-takashima-fuel-saving-weather-routing, 2018-grifoll-ship-weather-pathfinding}, however, within this paper we stick with simpler definitions.

This approach is similar to the one proposed in paper~\cite{2019-jeong-multi-criteria-route-planning}, however, we propose to use different definitions of the coefficients and to compare the route under consideration with the best possible route. We also assume that the initial criteria can be extended according to the needs of the particular voyage.

Since these criteria can be used to compare routes, it is possible to use evolutionary computations like genetic algorithms (\cite{1975-holland-adaptation-in-systems}), or any other optimisation method that relies on comparing solution candidates, to plan a route for a ship. As there are two criteria, route optimisation is a multicriteria optimisation task. Therefore, the optimisation algorithm may reach the Pareto-optimum. When Pareto-optimum is reached, it is not possible to improve one of the criteria in a way other than by reducing the value of another. In such cases, it is common practice to ask an expert to choose the best of the Pareto-optimal solutions. The other way is to ask an expert to select ``weights'' for criteria, replacing a multicriteria optimisation problem with unicriteria, where the solutions are compared using the weighted sum 
\begin{equation}\label{eq:overall-sum}
s = \sum_{i=1}^n w_i c_i
\end{equation}
where $c_i$ is the value of a particular criteria and $w_i$ is its weight. However, it is known that for some tasks even an expert may fail to choose these weights~\cite{2006-konak-multi-objective-opt-ga-tutorial}.

On the other hand, in case of safety vs. optimality dilemma, there are only two criteria and choosing which criteria is more important might be easier. We may consider safety to be more important than optimality. Indeed, safe but sub-optimal route is better than the route that is optimal, but unsafe. Therefore we may choose the best route in two steps: first, select a subset of Pareto-optimal solutions that have the highest safety criterion value and then select the one that has the highest optimality score. However, this approach is straightforward only when there are only two criteria. If there are more of them, ranking may be a more complicated task.

\section{Future route prediction}
Analysis of a route during the voyage helps determine whether the route will eventually be safe and optimal. However, it is more convenient to be able to determine the route's characteristics before it gets too late to change it. Therefore we need to check the route's safety and optimality for the rest of the route before the ship has passed it.

Consider a planned ship route from point $A$ to point $B$ that has $N$ waypoints and an actual route from point $A$ to point $B$. The actual route deviates from the planned one, therefore for each point $p_i$ of the planned route there is a corresponding point $p_i'$ of the actual route. Let the ship has arrived at the point $1'$. We can now compute characteristic coefficients for the actual route that consists of two subroutes: from point $A$ to point $1'$ and from point $1'$ to point $B$. Therefore we need to predict the subroute from the point $1'$, to the point $B$ in order to take into account the discrepancy between the actual route and the planned one. Here we assume that the ship keeps the discrepancy and performs exactly the same manoeuvre at the point $1'$ that has been planned for point $1$. This gives us the point $2'$, that corresponds to point $2$ of the planned route. 

\begin{wrapfigure}
  \centering
  \includegraphics[width=0.35\textwidth]{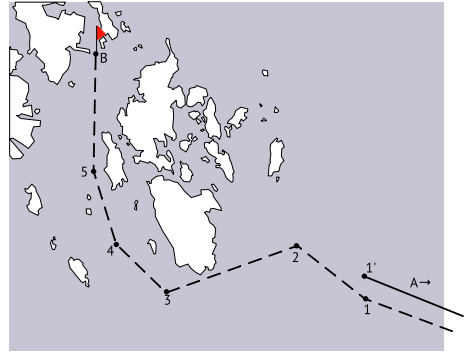}
  \caption{Sample route. The planned route is represented as the dashed line, the actual route is represented with the solid line.}
  \label{img:kotka-1}
\end{wrapfigure}

\begin{wrapfigure}  
\centering
    \includegraphics[width=0.35\textwidth]{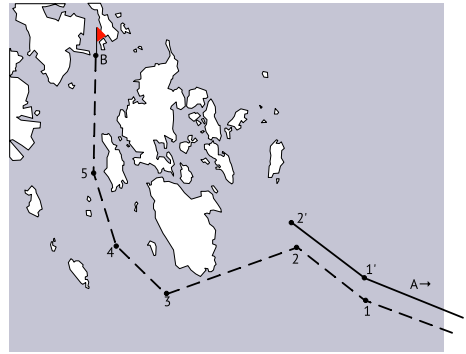}
    \caption{Ship route with the predicted edge.}
    \label{img:kotka-2}
\end{wrapfigure}

Consider the route depicted in the Figure~\ref{img:kotka-1}. As one can see, the actual route deviates from the planned one. This deviation means that the characteristic coefficients computed for the actual route differ from those computed for the planned one. However, it is not obvious whether this deviation affects the actual route in the long run. In order to determine these effects, we use route forecasting. Using the location of the point $1'$ and the manoeuvre the ship is expected to perform at the point $1$, we can predict the next edge of the actual route. Figure~\ref{img:kotka-2} shows the predicted edge from the point $1'$ to the point $2'$.

\begin{wrapfigure}
\centering
    \includegraphics[width=0.35\textwidth]{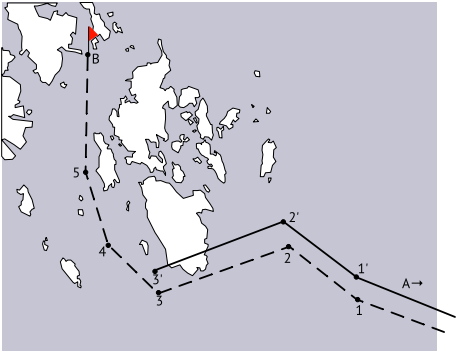}
    \caption{One more predicted edge. In this case, collision expected.}
    \label{img:kotka-3}
\end{wrapfigure}

We can repeat this action. Considering the expected location of the point $2'$, and using the manoeuvre planned for the point $2$, we can predict the location of the point $3'$ (Figure~\ref{img:kotka-3}). We continue this process until we get the location of the point $B'$.

When the prediction process is completed, we estimate the characteristic coefficients of the predicted route and compare them to the coefficients of the planned route. If the difference is not acceptable, a new route should be plotted starting from the current location of the ship. Figure~\ref{img:kotka-3} shows that the edge $2' \rightarrow 3'$ crosses the land, which means that the route is expected to be unsafe. The route forecast lets the ship detect potentially dangerous deviations and take necessary actions before it gets too late.



This method is based on the assumption that the ship keeps the already happened discrepancy between the planned and the actual route, which means it does not take any actions to resolve or deepen these discrepancies. However, the ship may take such actions, and perform at point $p_i'$ a manoeuvre that differs from the one planned for the point $p_i$. In this case the difference between the actual and the planned route changes. Therefore it may not be possible to determine that the discrepancy leads to a route that does not satisfy the applied safety or optimality requirements. To handle such situations, we recommend to analyze not only the route itself and its coefficients, but also the process of route development, as described in the next section.

\section{Ship route refinement}

If the ship keeps to its current movement regime then the difference between the actual route and the planned one will become more and more significant. We use route prediction to determine what the actual route will look like in the future. A predicted route consists of two parts: the part from the start waypoint till the current waypoint (the actual part), and the waypoint from the current waypoint till the destination waypoint (the predicted part). Therefore, each predicted route is a full route from the start to the destination. The more the ship has passed, the less the predicted part is.

It is known that analysis of a data represented as a set of numbers may be difficult for a human, especially in case of an extreme situation that requires fast reaction. To make it easier for the crew to analyse route quality we propose to describe the route as an image that encodes values of the coefficients~\cite{2011-nechaev-catastrophe-theory}. We call such an image a cognitive image, as it makes it easier to cognitively interpret the route quality measures. Through changes of that image the crew can see and analyse changes of the values of the coefficients and thus changes of the route being represented. Using such images one can easily compare two routes by comparing their images. The image must be easily understandable by the crew, which means that in case of any route change, the image must reflect it. If the route's safety measure changes, the safety coefficient also changes and the part of the image that encodes safety measure must also change. We think that a graph can be used as such an image. Such a graph should have five vertices -- one per coefficient and the additional central vertex that joins the vertices together into a single structure. The length of an edge that connects the certal vertex and the vertex that represents a coefficient encodes the value of this coefficient. The exact structure of the graph can be arbitrary, however the same structure must be used to represents all the routes being compared. We also recommend to draw a curve around the graph using its coefficient vertices as control points to make the image more representative. Figure~\ref{img:route-image} depicts the cognitive image for a route.

\begin{wraptable}
  \centering
  \includegraphics[width=3cm]{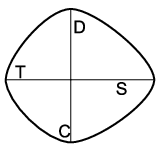}
  \caption{Cognitive image of a route.}
  \label{img:route-image}
\end{wraptable}

When a route changes, its coefficients also change, which makes the coefficient vertices change their locations, which, in turn, makes the overall shape of the image change. We consider such cognitive image representative since it allows comparing routes by comparing their complete images, without the need to analyze the coefficients. To properly examine the route we should compute the characteristic coefficients for the entire route, including both actual and predicted parts. Otherwise, the optimality coefficients will constantly grow with time, since the distance between the current location and the destination becomes shorter as the ship approaches it. This, in turn, makes distance, time and simplicity coefficients grow as the ship approaches the destination.

We assume that the cognitive image of a route either grows in size or stays the same with time. This means the actual route of the ship is either better than the planned one or, at least, is as good as it was. However, if the cognitive image shrinks in size, then the actual route tends to become worse than the planned one. If the route is expected to become unsafe or non-optimal to the extent that is not allowed, then a new route should be planned.

Consider a route that has been changing with time in a way, that is represented in table~\ref{table:route-evol-ok}.

\begin{wraptable}
    \caption{Characteristic coefficients of a route, successful case.}
    \centering
	
	\begin{tabular}{|p{1.5cm}|p{1cm}|p{1cm}|p{1cm}|p{1cm}|}
  \hline
  Waypoint  & S     & D    & T    & C    \\ \hline    
  1         & 0,53  & 0,45 & 0,35 & 0,1  \\ \hline    
  2         & 0,55  & 0,48 & 0,45 & 0,1  \\ \hline    
  3         & 0,65  & 0,58 & 0,6  & 0,25 \\ \hline    
  4         & 0,71  & 0,68 & 0,65 & 0,5  \\ \hline    
  5         & 1     & 0,85 & 0,78 & 0,75 \\ \hline    
  \end{tabular}
	\label{table:route-evol-ok}
\end{wraptable}

\begin{wrapfigure}
	\centering
	\includegraphics[width=0.5\textwidth]{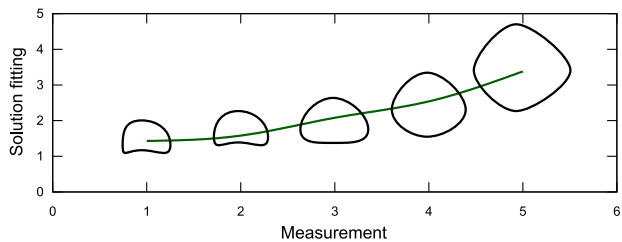}
	\captionof{figure}{The route quality grows, cognitive image goes larger in size, the route development process tends to produce a safe and optimal route.}
	\label{img:route-evol-ok}
\end{wrapfigure}

This table provides values of the characteristic coefficients of the actual route computed at each waypoint. Based on these values we can make cognitive images for the route at each of the waypoints. We can also compute overall route quality as the sum of the coefficients (equation~\ref{eq:overall-sum}). Figure~\ref{img:route-evol-ok} shows both overall route quality and cognitive images using coefficient values provided by the table~\ref{table:route-evol-ok}.

As one can see in the Figure~\ref{img:route-evol-ok}, the route development process tends to produce a safe and optimal route. Both the coefficients and the cognitive image tend to grow, which means that the development process stabilizes.

Table~\ref{table:route-evol-fail} provides characteristic coefficients of another route.

\begin{wraptable}
    \caption{Characteristic coefficients of a route, failure case.}
    \centering

	\begin{tabular}{|p{1.5cm}|p{1cm}|p{1cm}|p{1cm}|p{1cm}|}
  \hline
  Waypoint  & S     & D    & T    & C    \\ \hline  
  1         & 0,72  & 0,8  & 0,65 & 0,25 \\ \hline  
  2         & 0,7   & 0,8  & 0,58 & 0,25 \\ \hline  
  3         & 0,5   & 0,6  & 0,45 & 0,25 \\ \hline  
  4         & 0,65  & 0,55 & 0,38 & 0,25 \\ \hline  
  5         & 0,5   & 0,45 & 0,25 & 0,25 \\ \hline  
  \end{tabular}
	\label{table:route-evol-fail}
\end{wraptable}

\begin{wrapfigure}
	\centering
	\includegraphics[width=0.5\textwidth]{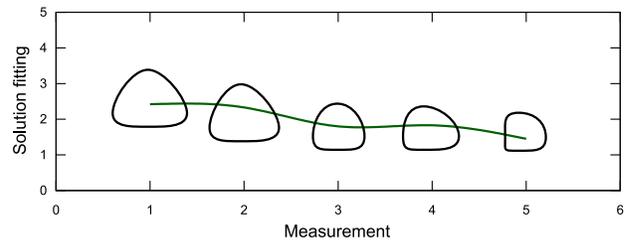}
	\captionof{figure}{The route quality decreases, the cognitive image shrinks, the route development process fails to produce a safe and optimal route.}
	\label{img:route-evol-fail}
\end{wrapfigure}

Figure~\ref{img:route-evol-fail} shows both overall route quality and cognitive images based on the coefficients provided in the table~\ref{table:route-evol-fail}. As one can see, the development process tends to fail to produce a safe and optimal route. The coefficient values decrease and the cognitive images shrink, which means that the process is unstable. In this case, it is recommended to plot a new route starting at the current ship's location.

\section{Accident analysis}
The method for route examination proposed in this paper can also be used to analyse a navigational accident. If the route development indicates that the route's quality decreases with time then the applied constraints may be violated, which means that the route development process is moving towards the ``possible accident'' state~\cite{2011-nechaev-catastrophe-theory}. In this case, a new route should be planned. If the development process stabilizes, yielding routes with similar coefficient values, that do not indicate violation of the constraints, then the discrepancy between the actual route and the planned one does not cause constraints violation~\cite{2011-nechaev-catastrophe-theory}. In this case, the route development process is moving towards the ``safe and optimal'' route state.

\begin{wrapfigure}
  \centering
  \includegraphics[width=0.5\textwidth]{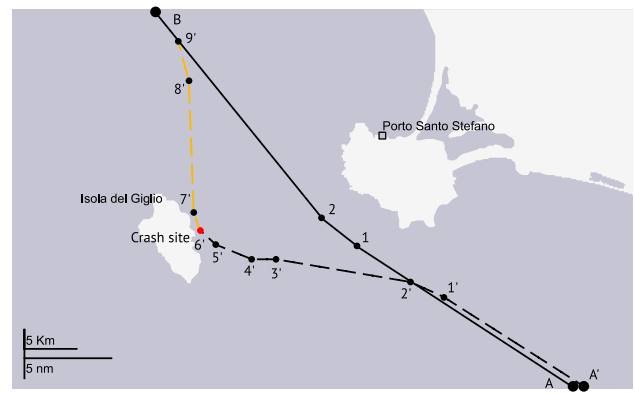}
  \captionof{figure}{Costa Concordia accident. The solid line is the route of the 6th of January, 2012, dashed line -- the route of the 13th of January, 2012, coloured line -- the possible return route.}
  \label{img:costa-concordia}
\end{wrapfigure}

In this paper we demonstrate the proposed method through the analysis of the well-known Costa Concordia accident, that happened on 13th of January, 2012. Figure~\ref{img:costa-concordia} compares the route of 6th of January, 2012 (the safe route, hereinafter ``the planned route'') to the route of the 13th of January, 2012 (the accident route, hereinafter ``the actual route'').

According to the official investigation report, the main reason of the accident was the deviation from the planned route, navigation too close to the shoreline and mistakes that have been made while planning the new, actual, route~\cite{2013-mit-costa-concordia-report}. 

In order to analyze a route we have to know the location of its waypoints. Since we do not know this both for the planned and actual routes, we have put it manually at the locations where the routes change the movement direction (Figure~\ref{img:costa-concordia}). According to the official investigation report, the crew was going to return to the planned route and continue the voyage~\cite{2013-mit-costa-concordia-report}. Figure~\ref{img:costa-concordia} shows the possible return route in colour. 

In order to compare the actual route (with the return part) to the planned route, we compute the characteristic coefficients for them. In this case the aim is not to examine the routes, but to compare them, therefore only the difference between the coefficient values is meaningful and not the values themselves.

The main difficulty here is the safety coefficient as it represents the propability of a collision with an obstacle. According to Figure~\ref{img:costa-concordia}, the planned route does not assume any collision, since it is located far from the shorelines. Therefore we may consider the safety coefficient for the planned route as $S = 1.0$ which means the ship is not at risk. In case of the actual route, it is known that the ship has already followed this dangerous route before~\cite{2015-bbc-costa-concordia-what-happened}, however, that time nothing has happened. Therefore we may assume the probability of an accident to be close to $0.5$, which makes the safety coefficient for the actual route be $S = 0.5$.

Table~\ref{table:costa-simple-compare} provides characteristic coefficient values for the routes. As one can see from this table, all coefficients computed for the actual route have smaller magnitude than those computed for the planned one. This means that the actual route is of lower quality compared to the planned one.

\begin{wraptable}
    \caption{Comparison of the characteristic coefficients for the routes of the 6th and 13th of January, 2012.}
    \centering

  \begin{tabular}{|p{2cm}|p{1.5cm}|p{1.5cm}|p{1.5cm}|}
  \hline
  Coefficient  & 6th January & 13th January & Difference \\ \hline
  S            & 1.0         & 0.5          & 0.5        \\ \hline
  D            & 0.9886      & 0.8667       & 0.1219     \\ \hline
  T            & 0.7879      & 0.6908       & 0.0971     \\ \hline
  C            & 0.5         & 0.1818       & 0.3182     \\ \hline
  \end{tabular}
  \label{table:costa-simple-compare}
\end{wraptable}

\begin{wrapfigure}
  \centering
  \includegraphics[width=3cm]{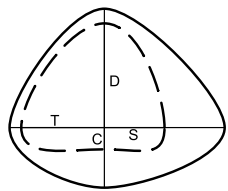}
  \captionof{figure}{Route cognitive images comparison. Solid line is for the 6th of January, 2012 route and the dashed line is for the 13th of January, 2012.}
  \label{img:costa-concordia-images}
\end{wrapfigure}

We can also compare the cognitive images for the routes. Figure~\ref{img:costa-concordia-images} shows the images both for the planned and actual routes. As one can see, these two images differ both in size and shape which represents the difference between the routes.

On the other hand, if the discrepancy was inadvertent, then the crew could have decided to return to the planned route earlier. In this case there could have been several possible return paths, depending on where the crew has decided to return. These possible return routes differ by the turn points, the points where the ship turns from the actual route in order to return to the planned one.

\begin{wrapfigure}
  \centering
  \includegraphics[width=0.5\textwidth]{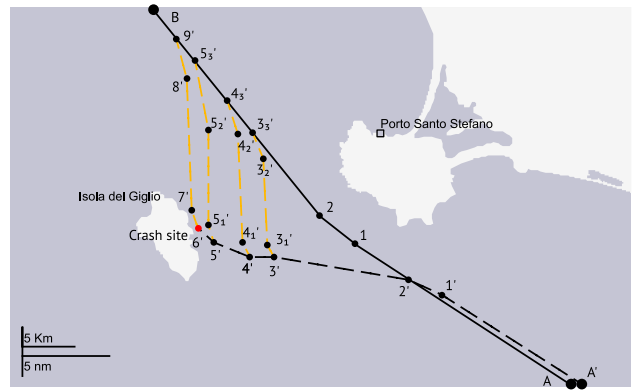}
  \captionof{figure}{Possible return routes depending on the actual route. The solid line is the expected route, dashed line is the actual route, coloured lines are the possible return routes.}
  \label{img:costa-concordia-routes}
\end{wrapfigure}

This can be considered as the analysis of predicted routes as described earlier, the difference is that the routes are not predicted but explicitly provided. Which turn point is chosen, depends on the moment of time when the decision is made. This makes the possible turn path change with time and leads to the emergence of the route development process. Figure~\ref{img:costa-concordia-routes} shows possible return routes for the intermediate waypoints of the actual route. The ship may follow one of these return paths to return to the planned route.

In order to analyse possible return paths, one needs to compute the characteristic coefficients for these routes. Table~\ref{table:costa-routes-compare} provides characteristic coefficient values of the routes from Point $A'$ to Point $B'$ depending on the turn point.

\begin{wraptable}
    \caption{Characteristic coefficients of the routes depending on the turn point.}
    \centering
    \begin{tabular}{|p{2cm}|l|l|l|l|}
    \hline
    \multicolumn{1}{|c|}{\multirow{2}{*}{Route}} & \multicolumn{4}{c|}{Coefficient} \\ \cline{2-5} 
    \multicolumn{1}{|c|}{}                       & S     & D      & T      & C      \\ \hline
    planned route                             & 1.0   & 0.9886 & 0.7879 & 0.5    \\ \hline
    From Point~3'                                & 0.875 & 0.9037 & 0.7202 & 0.25   \\ \hline
    From Point~4'                                & 0.75  & 0.8829 & 0.7036 & 0.2222 \\ \hline
    From Point~5'                                & 0.625 & 0.8672 & 0.6912 & 0.2    \\ \hline
    From point~6'                                & 0.5   & 0.8667 & 0.6908 & 0.1818 \\ \hline
    \end{tabular}
    \label{table:costa-routes-compare}
\end{wraptable}

The table shows how the route quality is changing while the ship is approaching the Point $6'$, where the accident has happened. As one can see, the greater the discrepancy between the planned route and the actual one, the smaller the values of the coefficients are. This means that the route development process is approaching the ``possible accident'' state~\cite{2011-nechaev-catastrophe-theory}. 

\begin{wrapfigure}
  \centering
  \includegraphics[width=0.5\textwidth]{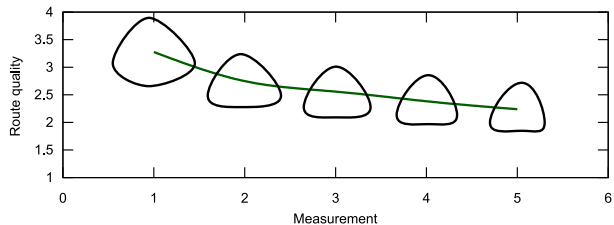}
  \captionof{figure}{The development of the route at the 13th of January, 2012 depending on the actual route.}
  \label{img:costa-concordia-images-change}
\end{wrapfigure}

Figure~\ref{img:costa-concordia-images-change} shows both overall route quality, computed as the sum of its coefficient values and the corresponding cognitive images. In this Figure, measurement~$1$ corresponds to the planned route, measurement~$2$ corresponds to the return route with the turn point~$3'$, measurement~$3$ -- to the return route with the turn point~$4'$, the measurement~$4$ -- to the return route with the turn point~$5'$ and the measurement~$5$ corresponds to the route though the point $6$, the crash point.

Therefore, the method described in this paper allows comparing routes. As a result, it is possible to analyse the effect of the crew's decisions on the route and its safety and optimality, including the decisions taken during the voyage.

It should be noted that in this section the coefficients were computed using several simplifications and assumptions and thus there might be minor inaccuracies. For example, the coefficient values do not take into account local currents that may affect ship movements. However, even in this case the actual route, that caused the accident, significantly differs from the planned one in terms of the coefficient values.

\section{Findings and conclusions}
This paper presents a novel approach to validation of a ship's route during the voyage based on route's characteristic coefficients. The method is based on the computation of several characteristic coefficients for the route and further analysis of the values. According to these coefficients, the crew can determine to what extent the route satisfies the applied safety and optimality requirements. The paper also presents a way to analyze the long-term effects of the discrepancies between the actual route and the planned one, using these coefficients. For the cases when the discrepancy changes with time, the paper presents the method to analyze the dynamic of the discrepancy. Using this dynamic the crew can analyze long-term effects of the current discrepancy between the routes using the history of the changes of the actual route.

The paper also describes the way the proposed method for estimating the route quality can be used while planning the ship's route using evolutionary computations or any other method that employs direct comparison of the candidate routes. Such methods are known to be used for route planning already~\cite{2015-simonsen-state-of-the-art-weather-routing}.

It is possible to introduce additional characteristic coefficients, in this case, the main idea of the method stays the same. Indeed, although the structure of the route's cognitive image changes due to additional coefficients, the method used to analyse the changes of the image stays the same. Such additional coefficients may represent ad-hoc requirements for the route that depend on the ship, the cargo, the laws and regulations or other factors.

We demonstrate the effectiveness of the method using the Costa Concordia accident. As one can see, the actual route significantly differs from the planned one in terms of the characteristic coefficient values and the difference grows as the ship approaches the crash point. 

We think that the proposed method can be used to improve shipping safety and efficiency. These coefficients can also be used to analyze a route not only during the planning process but also while investigating the reasons of a ship accident in order to determine whether the route affected the situation. The method can also be used to control navigation safety of an autonomous ship, because it uses formal criteria to measure route's properties and therefore can be implemented as a part of an on-board control system.

\end{multicols}

\providecommand*{\BibDash}{}
\urlstyle{rm}
\bibliography{local}{}
\bibliographystyle{plain}  
\urlstyle{tt}

\end{document}